\documentclass[runningheads]{llncs}

\usepackage[T1]{fontenc}
\usepackage{amsmath}
\usepackage{amssymb}
\usepackage{booktabs}
\usepackage{tabularx}
\usepackage{array}
\usepackage{graphicx}
\usepackage{xcolor}
\usepackage{url}
\newcolumntype{Y}{>{\centering\arraybackslash}X}

\newcommand{\method}{GRACE}
\newcommand{\ocav}{OCAV}

\newif\ifarxiv
\arxivtrue  

\begin{document}

\title{\method: Grounded Reasoning via Adapter Composition and Evidence-Aware Calibration for Educational Visual Question Answering}
\titlerunning{\method}
\author{Xinjin Li\inst{1}\textsuperscript{$\dagger$} \and
Yudi Xia\inst{2}\textsuperscript{$\dagger$} \and
Xi Zhao\inst{1}\textsuperscript{$\dagger$} \and
Yiliu Xu\inst{3} \and
Yining Liu\inst{4} \and
Cheng Lu\inst{5} \and
Yujian Long\inst{6} \and
Yu Ma\inst{3} \and
Jinghan Cao\inst{7} \and
Liang Fan\inst{8} \and
Yeyun Xu\inst{9}\textsuperscript{*}}
\authorrunning{Li et al.}
\institute{Columbia University, New York, NY, USA\\
\email{li.xinjin@columbia.edu; xz3124@tc.columbia.edu}
\and
Independent Researcher\\
\email{171250011xyd@gmail.com}
\and
Carnegie Mellon University, Pittsburgh, PA, USA\\
\email{yiliux@alumni.cmu.edu; yuma13926@gmail.com}
\and
University of California, Berkeley, Berkeley, CA, USA\\
\email{yiningliu610@berkeley.edu}
\and
Stevens Institute of Technology, Hoboken, NJ, USA\\
\email{chenglu@ieee.org}
\and
Georgetown University, Washington, DC, USA\\
\email{yl1140@georgetown.edu}
\and
San Francisco State University, San Francisco, CA, USA\\
\email{jcao3@alumni.sfsu.edu}
\and
Loughborough University, Loughborough, UK\\
\email{L.Fan@lboro.ac.uk}
\and
Texas A\&M University, College Station, TX, USA\\
\email{yeyun.xu1@gmail.com}\\[0.35em]
\textsuperscript{$\dagger$}These authors contributed equally to this work.\\
\textsuperscript{*}Corresponding author: Yeyun Xu}

\maketitle

\begin{abstract}
Educational visual question answering (VQA) requires models to solve curriculum-oriented multiple-choice questions using both language and visual evidence. Compared with conventional open-ended VQA, educational examples often include structured assessment metadata, diagrams or image contexts, and semantically close answer options, creating strong opportunities for question--option shortcuts. We develop and evaluate a parameter-efficient adaptation framework for a frozen multimodal large language model in this setting. We introduce \textbf{\method} (Grounded Reasoning via Adapter Composition and Evidence-Aware Calibration), a framework that uses each question's pedagogical state to specialize lightweight language and vision adaptation. The state combines inference-visible subject, grouped skill, grade, visual-context, question-intent, and option-structure cues. \method\ uses factor-specific prompts and lightweight visual adapters, then applies evidence-aware option calibration to score all candidates under a shared multimodal context. On ScienceQA, \method\ improves a shared-adapter baseline from 90.5\% to 93.1\% overall accuracy and from 88.7\% to 91.2\% on image-context questions. Removing pedagogical composition, option calibration, or the visual adapter reduces overall accuracy by 1.4, 1.0, and 1.5 points, respectively. These controlled results show that structured educational state is an effective routing signal for parameter-efficient multimodal adaptation.

\end{abstract}

\section{Introduction}

Multimodal large language models (MLLMs) have substantially improved visual question answering by combining image encoders with instruction-following language decoders~\cite{liu2023visual,li2023blip,dai2023instructblip,zhu2024minigpt}. Educational visual question answering, represented by ScienceQA~\cite{lu2022learn}, is more structured. Models must answer curriculum-oriented multiple-choice questions from a question stem, candidate options, optional diagrams or natural images, and assessment metadata. Correct prediction often requires deciding whether visual evidence is relevant, extracting the appropriate visual cue, connecting it to grade-appropriate concepts, and discriminating among plausible distractors. This makes educational VQA a useful setting for studying both multimodal reasoning and task-specific adaptation.

The adaptation problem is also different from generic visual instruction following. BLIP-2~\cite{li2023blip} demonstrates that frozen image and language models can be connected through a compact learned interface, while InstructBLIP~\cite{dai2023instructblip} and LLaVA~\cite{liu2023visual} show the effectiveness of instruction-aware multimodal adaptation and visual instruction tuning. ScienceQA adds explicit educational structure---subject, grade, skill, visual context, and option format---which we use to allocate limited trainable capacity on a per-example basis.

A second challenge is language shortcut learning. Educational questions often contain rich textual context, answer options, subject labels, and grade-level cues, so a model can sometimes answer from question--option statistics without relying sufficiently on the image. Aggregate accuracy alone does not reveal this behavior. At the same time, fully fine-tuning a large MLLM is expensive and unnecessary for a controlled adaptation study. We therefore keep the visual-transformer blocks and language decoder frozen, while training the multimodal connector and lightweight adaptation modules.

Our key observation is that educational benchmarks already expose structure that is useful for adaptation. Subject, grouped skill, grade, visual-context attributes, and question--option format cues are either part of the assessment instance or can be derived from fields visible before prediction. We combine these signals into a sample-specific \emph{pedagogical state}. Rather than assigning every example to one shared prompt or adapter, \method\ uses the state to compose a small bank of factor-specific prompt components and to gate residual visual adapters. This allows examples to share parameters along individual educational factors while still receiving different adaptation paths.

We additionally align the prediction rule with the multiple-choice task. All options for a question are scored under the same multimodal context. Our evidence-aware calibration is instantiated by \emph{Option-Calibrated Answer Verification} (\ocav), which combines length-normalized option likelihood with a lightweight multimodal verifier conditioned on the image, question, option, and pedagogical state. We additionally run a matched image-removal evaluation to quantify visual dependence of the resulting decision function.

We evaluate \method\ on ScienceQA using Qwen2.5-VL-7B-Instruct as the common backbone for the controlled comparisons. Relative to a shared adapter, \method\ improves overall accuracy from 90.5\% to 93.1\% and IMG accuracy from 88.7\% to 91.2\%. Relative to the frozen backbone, the gains are 5.6 and 6.3 points, respectively. Component ablations further show that pedagogical composition, option calibration, and visual adaptation each contribute to the final result.

Our contributions are as follows:
\begin{enumerate}
    \item We formulate educational VQA as a metadata-structured parameter-efficient adaptation problem in which inference-visible educational factors determine how limited trainable capacity is allocated across heterogeneous questions.
    \item We introduce pedagogy-aware adapter composition, which factorizes subject, skill, grade, visual context, question intent, and option structure into a sample-specific state that composes language prefixes and gates visual residual adapters.
    \item We introduce evidence-aware option calibration through \ocav, which scores the supplied candidates directly under a shared multimodal context and trains the correct option to outrank its distractors.
    \item Controlled ScienceQA experiments compare a frozen backbone, LoRA, a shared adapter, the full \method\ configuration, and core ablations. \method\ reaches 93.1\% overall accuracy and outperforms all controlled parameter-efficient baselines.
\end{enumerate}

\section{Related Work}

\subsection{Educational VQA}
ScienceQA~\cite{lu2022learn} introduced a large-scale multimodal multiple-choice benchmark for science education, where examples can include questions, answer options, images, subject and grade metadata, lectures, and explanations. Unlike the open-ended answer generation emphasized in conventional VQA~\cite{antol2015vqa}, ScienceQA makes option comparison explicit and couples it with curriculum metadata and educational visual reasoning. Rationale-augmented approaches such as Multimodal-CoT~\cite{zhang2024multimodal} use generated or annotated reasoning chains to improve answer inference. Our study targets a different axis: how inference-visible educational structure can route parameter-efficient adaptation. Lectures and explanations are not used to construct the routing state or answer-prediction input.

\subsection{Parameter-Efficient Adaptation of Vision-Language Models}
Parameter-efficient tuning keeps most pretrained weights fixed while optimizing compact modules such as prefix tuning~\cite{li2021prefix}, prompt tuning~\cite{lester2021power}, adapters~\cite{houlsby2019parameter,chen2022adaptformer}, and LoRA~\cite{hu2022lora}. Multimodal extensions adapt vision-language representations with limited trainable capacity~\cite{zhang2024llamaadapter,yang2024mma,luo2023cheap}. Importantly, prior multimodal PEFT is not restricted to a single shared update: LaVIN's Mixture-of-Modality Adaptation (MMA)~\cite{luo2023cheap}, for example, includes modality-aware routing between single- and multimodal instructions. \method\ differs in the routing signal and composition granularity. It factorizes multiple educational properties---subject, skill, grade, visual context, question intent, and option structure---and composes several factor-specific components for each assessment item rather than routing primarily by modality.

\ifarxiv
Related adaptive designs have also been explored in multi-agent learning, including Bayesian critique-tuning with adaptive pressure for multi-intersection traffic signal control~\cite{11059994} and low-frequency truncation for adaptive context-length optimization in multi-agent reinforcement learning~\cite{duanadaptive}. MAVEN-T further provides an efficiency-oriented example by using reinforced heterogeneous distillation for real-time multi-agent trajectory prediction~\cite{duan2026maven}.
\fi

\subsection{Visual Reliance and Multiple-Choice Scoring}
VQA systems can exploit dataset priors and language statistics even when visual evidence is weak~\cite{goyal2017making,agrawal2018dont}. For image-context educational questions, matched image interventions provide a direct way to measure how much the decision changes when visual content is removed. Multiple-choice evaluation also benefits from scoring the supplied candidates directly rather than relying on unconstrained answer strings; prior work on language-model evaluation and calibration shows that option and prompt biases can materially affect finite-choice decisions~\cite{zhao2021calibrate}. \method\ addresses these issues separately: \ocav\ aligns training and inference with the candidate set, while the image-removal analysis measures visual dependence.

\section{Method}

\subsection{Problem Setup \& Overview}
We consider multiple-choice educational VQA. Let
\[
    e_i=(I_i,q_i,O_i,y_i,m_i)
\]
denote an example, where $I_i$ is an image or visual context, $q_i$ is the question, $O_i=\{o_{i1},\ldots,o_{iK}\}$ is the candidate set, $y_i$ is the correct option index, and $m_i$ is benchmark metadata. The backbone is Qwen2.5-VL-7B-Instruct~\cite{qwen25vl}. Its visual-transformer blocks and language decoder remain frozen; the visual-to-language multimodal connector and lightweight modules $\Theta_A$ are trainable. At inference, the model receives only the image or visual context, question, candidate options, and inference-visible metadata. Gold answers, solution fields, lectures, explanations, and post-answer labels are never used to construct the adaptation state or prediction input.

The model predicts
\[
    \hat{y}_i=\arg\max_j S_{ij},
\]
where $S_{ij}$ is the evidence-aware calibrated score assigned to option $o_{ij}$.

\begin{figure}[t]
\centering
\includegraphics[width=\linewidth]{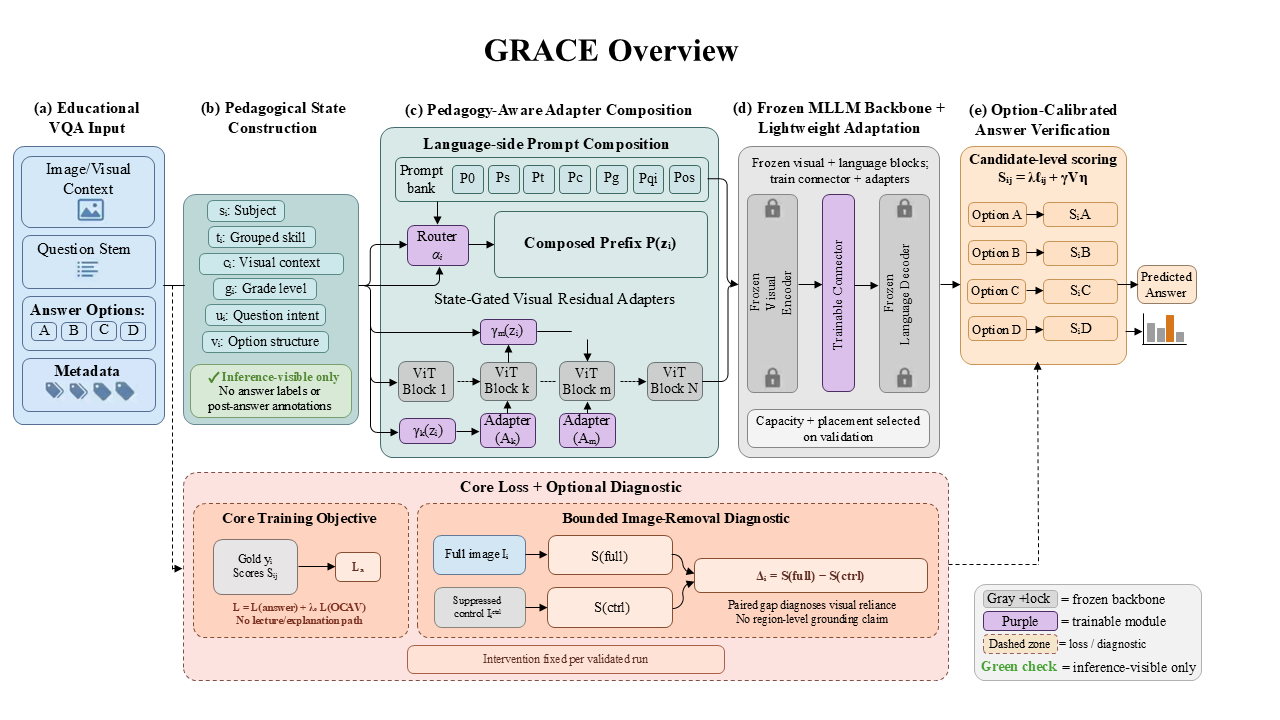}
\caption{Overview of \method. Inference-visible educational fields define a pedagogical state that composes factor-specific language prefixes and gates lightweight visual residual adapters. The visual-transformer blocks and language decoder are frozen, while the multimodal connector and lightweight adaptation modules are trained. Candidate options are scored under the same multimodal context by evidence-aware option calibration.}
\label{fig:overview}
\end{figure}

Figure~\ref{fig:overview} summarizes the pipeline. A pedagogical state first encodes the educational characteristics of the example. The state controls language-side prefix composition and visual-side residual adaptation. The adapted representations are then used to score every answer option under a common context.

\subsection{Pedagogy-Aware Adapter Composition}
Given an example, \method\ constructs a pedagogical state
\[
    z_i=\{s_i,t_i,c_i,g_i,u_i,v_i\},
\]
where $s_i$ is subject or topic, $t_i$ is a grouped pedagogical skill category, $c_i$ is visual context, $g_i$ is grade level, $u_i$ is question-intent category, and $v_i$ is an option-structure descriptor. The components $(s_i,t_i,c_i,g_i)$ come from benchmark metadata, while $(u_i,v_i)$ are computed from inference-visible question and option fields.

We maintain the factorized prompt bank
\[
\mathcal{P}=\{P_0,P^s,P^t,P^c,P^g,P^{\mathrm{qi}},P^{\mathrm{os}}\},
\]
where $P_0$ is shared and the remaining banks correspond to the observed state factors. The final implementation contains 93 prompt components, each with prefix length $l=10$. For state $z_i$, the composed prefix is
\[
\begin{aligned}
P(z_i)=P_0
&+\alpha_s P^s_{s_i}+\alpha_t P^t_{t_i}+\alpha_c P^c_{c_i}
+\alpha_g P^g_{g_i}\\
&+\alpha_{qi}P^{\mathrm{qi}}_{u_i}+\alpha_{os}P^{\mathrm{os}}_{v_i}.
\end{aligned}
\]
The coefficients $\alpha$ are produced by a small gating network over embeddings of the active metadata categories and question--option descriptors; missing fields use an explicit unknown-category embedding. This factorization lets related questions share individual subject, skill, grade, or visual-context components without requiring a separate expert for every joint state.

If factor $f$ has $n_f$ observed categories, a monolithic expert for every joint state would scale approximately with $\prod_f n_f$, whereas the factorized bank scales with $1+\sum_f n_f$. The latter is better matched to educational datasets in which many joint combinations are sparse.

\subsection{State-Conditioned Visual Adaptation}
Language-side composition alone does not specialize visual feature processing. We therefore insert two-layer bottleneck adapters with bottleneck dimension 256 at visual-transformer blocks 4 and 8. For hidden representation $h_i^{(\ell)}$ at an adapted layer $\ell$, the residual adapter is
\[
    A_\ell(h)=W^{(\ell)}_{\mathrm{up}}\,\sigma\!\left(W^{(\ell)}_{\mathrm{down}}h\right),
\]
and the state-conditioned update is
\[
    \tilde{h}_i^{(\ell)}=h_i^{(\ell)}+a_i^{(\ell)}A_\ell\!\left(h_i^{(\ell)}\right),
    \qquad a_i^{(\ell)}=G_\ell(z_i)\in[0,1].
\]
The gate $G_\ell$ uses the same pedagogical state as the prompt composer. Thus, the language and visual paths are conditioned on a common educational description while operating on different representations.

\subsection{Evidence-Aware Option Calibration}
Free-form generation is poorly matched to a finite candidate set because formatting and verbose output can introduce avoidable answer-normalization errors. We instead score every supplied option under the same multimodal context. For option $o_{ij}$, we compute the length-normalized conditional log-likelihood
\[
    \ell_{ij}=\frac{1}{|o_{ij}|}\sum_{t=1}^{|o_{ij}|}
    \log p_{\theta,\Theta_A}(o_{ij,t}\mid I_i,q_i,O_i,z_i,o_{ij,<t}).
\]
Length normalization removes the mechanical accumulation advantage of shorter or longer option strings. A lightweight verifier $V_\eta$ then evaluates the candidate jointly with the image, question, and pedagogical state. The final score is
\[
    S_{ij}=\lambda\ell_{ij}+\gamma V_\eta(I_i,q_i,o_{ij},z_i).
\]
We use $\lambda=1$ and $\gamma=1$ in the final configuration. We refer to this mechanism as \emph{Evidence-Aware Calibration}: the finite answer set is calibrated using both conditional option likelihood and a verifier that has access to the same multimodal evidence used by the predictor. The verifier treats the correct option as positive and the remaining options as negatives. \ocav\ is the listwise training implementation of this calibration mechanism.

The option-calibrated objective is
\[
    \mathcal{L}_{ocav}=-\log
    \frac{\exp(S_{iy_i})}{\sum_{j=1}^{K}\exp(S_{ij})}.
\]
This directly trains the correct option to outrank all distractors in the same question.

\subsection{Training Objective and Visual-Dependence Evaluation}
The answer loss is
\[
    \mathcal{L}_{ans}=-\log p_{\theta,\Theta_A}(o_{iy_i}\mid I_i,q_i,O_i,z_i),
\]
and the final training objective is
\[
    \mathcal{L}=\mathcal{L}_{ans}+\lambda_o\mathcal{L}_{ocav},
\]
with $\lambda_o=1.0$ in the final configuration.

For visual-dependence evaluation on image-available examples, we compare the original image with an all-gray control image of the same size, processed through the identical model and preprocessing path. We run the same option-scoring rule in both conditions and report full-image accuracy, blank-image accuracy, and their difference on the image subset. This matched intervention changes only visual content and is used for evaluation; it is not an additional inference input.

\subsection{Parameter Efficiency}
We train the multimodal connector together with the grouped prompt bank, metadata gates, visual adapters, and option verifier. These components total approximately 49.3M trainable parameters, about 0.7\% of the 7B checkpoint. The visual-transformer blocks and language decoder remain frozen.

\section{Experiments}

\subsection{Experimental Setup}

\paragraph{Dataset and protocol.}
We evaluate on \textbf{ScienceQA}~\cite{lu2022learn}, which contains 21,208 multimodal multiple-choice questions with official train, validation, and test splits of 12,726, 4,241, and 4,241 examples. We use the official splits. The pedagogical state is constructed only from information visible before answer prediction: benchmark-provided subject, grade, topic/category/skill, and visual-context metadata, together with question-intent and option-structure descriptors derived from the question and candidate options. No gold answer, answer index, validation/test label, solution field, lecture, explanation, or other post-answer annotation is used by the state parser, prompt composer, visual-adapter gate, option verifier, or any answer-prediction component at validation or test time. Hyperparameters are selected on the validation split, and the test split is evaluated after model selection. We use no additional ScienceQA training examples, synthetic questions, evidence boxes, or manually added annotations.

\paragraph{Controlled comparison setting.}
All author-run comparisons use Qwen2.5-VL-7B-Instruct~\cite{qwen25vl}, the same ScienceQA splits, processor, prompt template, and evaluation code. The controlled baselines are the frozen backbone, LoRA~\cite{hu2022lora}, and our shared adapter baseline implemented as a single Houlsby-style adapter path without pedagogical composition~\cite{houlsby2019parameter}. Published methods in Table~\ref{tab:main} are included for task context; the controlled conclusions are drawn from the common-backbone author-run rows.

\paragraph{Backbone and adapters.}
The Qwen2.5-VL visual-transformer blocks and language decoder are frozen, while the visual-to-language connector is trainable. The grouped prompt bank contains 93 length-10 components. Visual adapters are two-layer bottleneck MLPs with bottleneck dimension 256 and are inserted at visual-transformer blocks 4 and 8. The full model contains approximately 49.3M trainable parameters.

\paragraph{Training details.}
We optimize the trainable connector, prompt bank, metadata gates, visual adapters, and option verifier with AdamW~\cite{loshchilov2019decoupled}, using learning rate $2\times10^{-4}$, batch size 16, weight decay 0.01, and a maximum of 5 epochs. Early stopping is based on validation accuracy. All controlled baselines use the same Qwen2.5-VL processor settings, prompt formatting, and evaluation script.

\paragraph{Metrics.}
The main metric is multiple-choice accuracy on the ScienceQA test set. We report Overall accuracy, image-available accuracy (IMG), textual-context accuracy (TXT), and grades 1--6 accuracy (G1--6). IMG and TXT are benchmark-defined slices and are not necessarily disjoint. For the visual-dependence analysis, we also report full-image and blank-image accuracy on the image subset.

\subsection{Main Comparison}

\begin{table}[t]
\centering
\small
\caption{Main results on ScienceQA. Published methods are reference rows; the Qwen2.5-VL rows are controlled author-run comparisons under the same backbone and evaluation protocol.}
\label{tab:main}
\setlength{\tabcolsep}{4.2pt}
\begin{tabular}{lccccc}
\toprule
\textbf{Method} & \textbf{Backbone} & \textbf{TXT} & \textbf{IMG} & \textbf{G1--6} & \textbf{Overall} \\
\midrule
\multicolumn{6}{l}{\emph{Published reference results}}\\
CoT GPT-3 (ALE)~\cite{lu2022learn}
& GPT-3 & 74.68 & 67.43 & 78.23 & 75.17 \\
LLaMA-Adapter~\cite{zhang2024llamaadapter}
& LLaMA-7B & 83.72 & 80.32 & 85.83 & 85.19 \\
Multimodal-CoT$_{\mathrm{Large}}$~\cite{zhang2024multimodal}
& T5-large & 90.13 & 88.25 & 91.12 & 90.45 \\
LLaVA+GPT-4 (judge)~\cite{liu2023visual}
& Vicuna-13B & 90.62 & 88.99 & 92.73 & 92.53 \\
\midrule
\multicolumn{6}{l}{\emph{Controlled Qwen2.5-VL-7B comparisons}}\\
Qwen2.5-VL-7B frozen~\cite{qwen25vl}
& Qwen2.5-VL-7B & 88.0 & 84.9 & 88.8 & 87.5 \\
LoRA~\cite{hu2022lora}
& Qwen2.5-VL-7B & 90.5 & 87.6 & 91.0 & 90.0 \\
Shared adapter (ours)~\cite{houlsby2019parameter}
& Qwen2.5-VL-7B & 91.1 & 88.7 & 91.7 & 90.5 \\
\textbf{\method}
& Qwen2.5-VL-7B & \textbf{93.7} & \textbf{91.2} & \textbf{93.8} & \textbf{93.1} \\
\bottomrule
\end{tabular}
\end{table}

Table~\ref{tab:main} shows a consistent advantage for \method\ under the common Qwen2.5-VL backbone. Compared with the frozen model, \method\ improves Overall accuracy from 87.5\% to 93.1\% (+5.6 points) and IMG accuracy from 84.9\% to 91.2\% (+6.3 points). The stricter comparison is the shared adapter, which uses the same backbone and a comparable lightweight adaptation path without pedagogical composition. \method\ improves that baseline by 2.6 points Overall, 2.5 points on IMG, 2.6 points on TXT, and 2.1 points on G1--6. It also exceeds LoRA by 3.1 points Overall and 3.6 points on IMG. The published rows use different backbones and training protocols, so we use them as reference points rather than as controlled estimates of the \method\ effect.

\subsection{Core Ablation Study}

\begin{table}[t]
\centering
\small
\caption{Core ablations on ScienceQA. Drop is the Overall accuracy decrease relative to the full \method\ configuration.}
\label{tab:ablation}
\begin{tabular}{lcccc}
\toprule
\textbf{Variant} & \textbf{IMG} & \textbf{TXT} & \textbf{Overall} & \textbf{Drop} \\
\midrule
\textbf{\method} & \textbf{91.2} & \textbf{93.7} & \textbf{93.1} & -- \\
w/o pedagogical composition & 89.2 & 92.0 & 91.7 & 1.4 \\
w/o option calibration & 90.1 & 92.8 & 92.1 & 1.0 \\
w/o visual adapter & 88.8 & 92.4 & 91.6 & 1.5 \\
\bottomrule
\end{tabular}
\end{table}

Table~\ref{tab:ablation} isolates the three mechanisms that define the current architecture. Removing the visual adapter causes the largest Overall decrease (1.5 points) and reduces IMG accuracy from 91.2\% to 88.8\%, confirming that the trainable visual residual path is important for image-context questions. Removing pedagogical composition reduces Overall accuracy by 1.4 points and lowers both IMG and TXT accuracy, showing that state-conditioned routing contributes beyond a single shared update. Removing evidence-aware option calibration reduces Overall accuracy by 1.0 point, demonstrating that candidate-level ranking contributes on top of representation adaptation.

\subsection{Visual-Dependence Analysis}
\begin{table}[t]
\centering
\small
\caption{Visual-dependence evaluation on the ScienceQA image subset. The blank-image condition replaces the image with an all-gray input of the same size while preserving the model and preprocessing path.}
\label{tab:visualdep}
\begin{tabular}{lccc}
\toprule
\textbf{Method} & \textbf{Full image} & \textbf{Blank image} & \textbf{Gap} \\
\midrule
Shared adapter (ours) & 88.7 & 80.9 & 7.8 \\
\textbf{\method} & \textbf{91.2} & 78.4 & \textbf{12.8} \\
\bottomrule
\end{tabular}
\end{table}

The matched intervention in Table~\ref{tab:visualdep} shows that \method\ both improves full-image accuracy and increases the full--blank gap from 7.8 to 12.8 points relative to the shared adapter. The larger paired degradation after visual content is removed shows that the resulting decision function relies more strongly on image content for image-context questions.

\section{Conclusion}

We presented \method, a parameter-efficient framework for educational VQA that uses inference-visible pedagogical state to compose language prefixes and gate visual residual adapters while keeping the large visual-transformer blocks and language decoder frozen. Evidence-aware option calibration aligns the model with the finite multiple-choice decision, and the common-backbone ScienceQA experiments show that the resulting architecture improves the shared-adapter baseline from 90.5\% to 93.1\% Overall and from 88.7\% to 91.2\% on IMG. Core ablations show independent performance losses when pedagogical composition, option calibration, or the visual adapter is removed, and the matched blank-image analysis shows stronger visual dependence than the shared adapter. Together, these results establish structured educational state as an effective routing signal for parameter-efficient multimodal adaptation on ScienceQA.

\paragraph{Limitations.}
The current evaluation is centered on ScienceQA and assumes that inference-visible educational metadata such as subject and grade is available. The blank-image intervention measures dependence on visual content but does not identify which image region caused a prediction. Evaluation on additional educational benchmarks, backbones, and metadata regimes is needed to determine how broadly the same routing design transfers.

\subsubsection*{Disclosure of Interests.}
The authors have no competing interests to declare that are relevant to the content of this article.

\ifarxiv
\nocite{zhang2026performance,liang2026hybridmetalearnersestimatingheterogeneous,chen2025autoneuralcodesigningvisionlanguagemodels,liu2026megaslideditmemorycentricadaptationdeformable,11622841}
\fi

{\small
\bibliographystyle{splncs04}
\ifarxiv
\bibliography{ref,ref_arxiv}
\else
\bibliography{ref}
\fi
}

\end{document}